\documentclass[superscriptaddress,amsmath,amssymb,aps,prx,longbibliography]{revtex4-2}

\usepackage{graphics,epsf}

\usepackage{graphicx}
\usepackage{dcolumn}
\usepackage{bm}
\usepackage[colorlinks=true, allcolors=blue]{hyperref}

\usepackage[capitalise]{cleveref}
\usepackage{amsmath}
\usepackage{amssymb}
\usepackage{amsfonts}

\usepackage{epstopdf}
\usepackage{amsbsy}
\usepackage[LGR,T1]{fontenc}
\usepackage{lmodern}
\usepackage{textgreek}

\usepackage{bbm}
\usepackage{braket}
\usepackage{xcolor}
\usepackage{float}
\usepackage[utf8]{inputenc}

\newcommand{\angstrom}{\mbox{\normalfont\AA}}

\renewcommand{\approx}{\simeq}

\begin{document}

\title{Emergence of a Fermi-surface in the current-driven hidden state of 1T-TaS$_2$ }

\author{Yuval Nitzav}
\affiliation{Department of Physics, Technion, Haifa, 3200003, Israel}
	\author{Roni Anna Gofman Kiassi}
    \affiliation{Department of Physics, Technion, Haifa, 3200003, Israel}
	\author{Ilay Mangel}
   \affiliation{Department of Physics, Technion, Haifa, 3200003, Israel}
 \author{Abigail Dishi}
   \affiliation{Department of Physics, Technion, Haifa, 3200003, Israel}
 \author{Nitzan Ragoler}
    \affiliation{Department of Physics, Technion, Haifa, 3200003, Israel}
\author{Sajilesh K.P.}
   \affiliation{Department of Physics, Technion, Haifa, 3200003, Israel}
 \author{Yaron Jarach}
 \affiliation{Department of Physics, Technion, Haifa, 3200003, Israel}

    \author{Alex Louat}
       \affiliation{Diamond Light Source, Harwell Science and Innovation Campus, Didcot, OX11 0DE, UK.}

    \author{Matthew D. Watson}
       \affiliation{Diamond Light Source, Harwell Science and Innovation Campus, Didcot, OX11 0DE, UK.}

    \author{Cephise Cacho}
   \affiliation{Diamond Light Source, Harwell Science and Innovation Campus, Didcot, OX11 0DE, UK.}
 \author{Irena Feldman}
\affiliation{Department of Physics, Technion, Haifa, 3200003, Israel}
\author{Amit Kanigel}
\affiliation{Department of Physics, Technion, Haifa, 3200003, Israel}

	\begin{abstract} The origin of the insulating state in 1T-TaS$_2$ has long been a subject of debate. A short current pulse transforms this insulating state into a metastable metallic phase. Using micro-ARPES, we investigate the electronic structure of this phase and uncover spatially dependent modifications caused by the current pulse. In some regions of the sample, a Fermi surface emerges, while others remain gapped. Detailed band structure analysis reveals that the metallic regions exhibit an electronic structure similar to that observed in the high-temperature phase of 1T-TaS$_2$, characterized by suppressed energy gaps and bands crossing the Fermi level. Furthermore, the metallic and insulating regions display distinct dispersions along the out-of-plane direction. These observations suggest a scenario in which the current pulse breaks the star-of-David dimers characteristic of the insulating phase, implicating these dimers as the likely origin of the insulating behavior in 1T-TaS$_2$.
    \end{abstract}


\maketitle


\noindent

Metastable and hidden states have emerged as key areas of interest in condensed matter physics, offering insights into novel phases of matter and exotic physical properties. They also present promising avenues for technological applications. A metastable state occurs when a system becomes trapped in a local energy minimum rather than reaching its global ground state. These states typically arise from competing orders\cite{oike2018kinetic,han2015exploration,madan2018nonequilibrium}, topological defects\cite{gerasimenko_intertwined_2019}, structural configurations\cite{frigge2017optically}, and other degrees of freedom that create a complex energy landscape with multiple stable configurations. Some metastable states can be accessed thermally, for example, by rapidly cooling a material from elevated temperatures~\cite{DISALVO19731357_4Hb,sung_two-dimensional_2022}. Distinctly,  hidden states represent a subset of metastable states that require non-thermal stimuli. These hidden configurations can only be reached through external drivers like ultrafast laser pulses or strong electric fields.~\cite{budden_evidence_2021,zhang_cooperative_2016,zong_ultrafast_2018,de2021colloquium}.

The study of these metastable states provides valuable insights into material properties and underlying physical mechanisms. By examining metastable configurations, we can better understand fundamental interactions in materials, for example electron-phonon coupling\cite{shi2019ultrafast} and magnetic-lattice interactions\cite{zhang_cooperative_2016}.



One such hidden phase is the metallic state induced at low temperatures in 1T-TaS$_2$ by various stimuli, including a short light pulse~\cite{stojchevska_ultrafast_2014}, charge transfer~\cite{ma2016metallic}, or a short electric current pulse~\cite{vaskivskyi_fast_2016}. Regardless of the perturbation method, the hidden state exhibits a similar critical temperature~\cite{venturini2022ultraefficient} and relaxation time scales~\cite{vaskivskyi_controlling_2015}. Following the short pulse, the resistance of the samples is reduced by up to several orders of magnitude. This hidden metallic state has been shown to persist for hours and is fully reversible either when the sample is heated above a certain temperature~\cite{vaskivskyi_controlling_2015} or subjected to an erasing pulse~\cite{stojchevska_ultrafast_2014}.

 Due to the fast and controllable nature of the transition to the hidden state~\cite{ARPES_MAKLAR_2023}, it has been suggested for various applications, including programmable light manipulation~\cite{vaskivskyi2024_programmable} and charge configuration memory (CCM) devices~\cite{mihailovic2021ultrafast,mraz2022charge,venturini2022ultraefficient,vaskivskyi_fast_2016}.

1T-TaS$_2$ exhibits an insulating ground state, whose origin remains unknown. The electronic-structure  is governed by a charge density wave (CDW) instability~\cite{bayliss_thermal_1984,rossnagel_origin_2011}. The main motif of the CDW is a rearrangement of every 13 Ta atoms in a star-of-David cluster where 12 Ta atoms are displaced towards the middle 13th atom. 
Below $352~K$ domains of star-of-David super-cells start to form. The domains are separated by regions with no CDW and form a hexagonal lattice which is nearly commensurate (NC) with the atomic lattice~\cite{wu1989hexagonal,thomson_scanning_1994,wu_direct_1990}.
Below $183~K$ all the star-of-David domains merge to create a commensurate CDW (C-CDW) state. The C-CDW vectors are rotated by $\pm$13.9 degrees from the atomic lattice vectors and form a new $\sqrt{13}\times\sqrt{13}$  sized unit cell. The orientation of the CDW vectors breaks mirror symmetry, and gives rise to two degenerate configurations (see Fig.~\ref{fig:Fermi_surface}E) with a distinct chirality. The sample's chirality is established during the formation of the CDW domains in the NC-CDW phase and persists into the C-CDW state~\cite{yang_chirality_2022}, where these chiral domains can extend to millimeter-scale sizes~\cite{liu2023electrical}.

The phase transition from NC-CDW to C-CDW is marked by an abrupt increase in resistivity, followed by strong insulating behavior at lower temperatures. In the C-CDW state, the valence band splits into seven narrow bands, six of which are fully occupied, while the topmost band is half-filled. The presence of unpaired electrons suggests that the C-CDW alone cannot fully explain the observed insulating state~\cite{tosatti_nature_1976,law20171t}.

It has been proposed that the narrow half-filled band in 1T-TaS$_2$ undergoes a Mott transition to an insulating state~\cite{Effects_of_stacking_Wu_2022,fazekas_charge_1980}. In recent years, it has become evident that the out-of-plane stacking of star-of-David clusters plays a significant role in determining the electronic properties of 1T-TaS$_2$~\cite{butler_mottness_2020,lee2019origin}. Evidence for out-of-plane dimerization is observed during the transition to the C-CDW state~\cite{stahl_collapse_2020}, resulting in an even number of electrons per unit cell and leading to insulating behavior~\cite{Stacking-driven_gap_Ritschel_2018}. STM measurement reveal that the monolayer 1T-TaS$_2$ is insulating\cite{vavno2021artificial}, consistent with Mott insulating behavior in the absence of inter-layer coupling. However, it remains unclear whether this behavior persists in the bulk, where out-of-plane interactions play a critical role. Understanding the nature of the metallic hidden state may provide insights into the origin of the insulating ground state.
 
In this paper, we use spatially resolved ARPES to track the evolution of the electronic structure of 1T-TaS$_2$ after a short current pulse. Our main results are summarized in Fig.~\ref{fig:fig1}. Following a short current pulse, the resistance of the sample at low temperatures decreases by a factor of about four (Fig.~\ref{fig:fig1}A). In this state, we observe a non-uniform electronic structure, with different parts of the sample exhibiting distinct behaviors (Fig.~\ref{fig:fig1}B). Most of the sample remains in the C-CDW state, characterized by a full gap and a clear band folding due to the CDW, both in-plane and out-of-plane. However, other regions, which contribute to the enhanced conductance in the hidden state, display a much weaker CDW effect and have a clear Fermi surface (Fig.~\ref{fig:fig1}C). The ARPES spectrum in these parts has some similarities with the NC-CDW spectrum although STM measurements find a completely different domain morphology in the two states~\cite{gerasimenko_intertwined_2019}. 
\section*{Results}
\subsection*{Experimental setup}
\begin{figure}[h]
    \centering
    \includegraphics[width=0.8\textwidth]{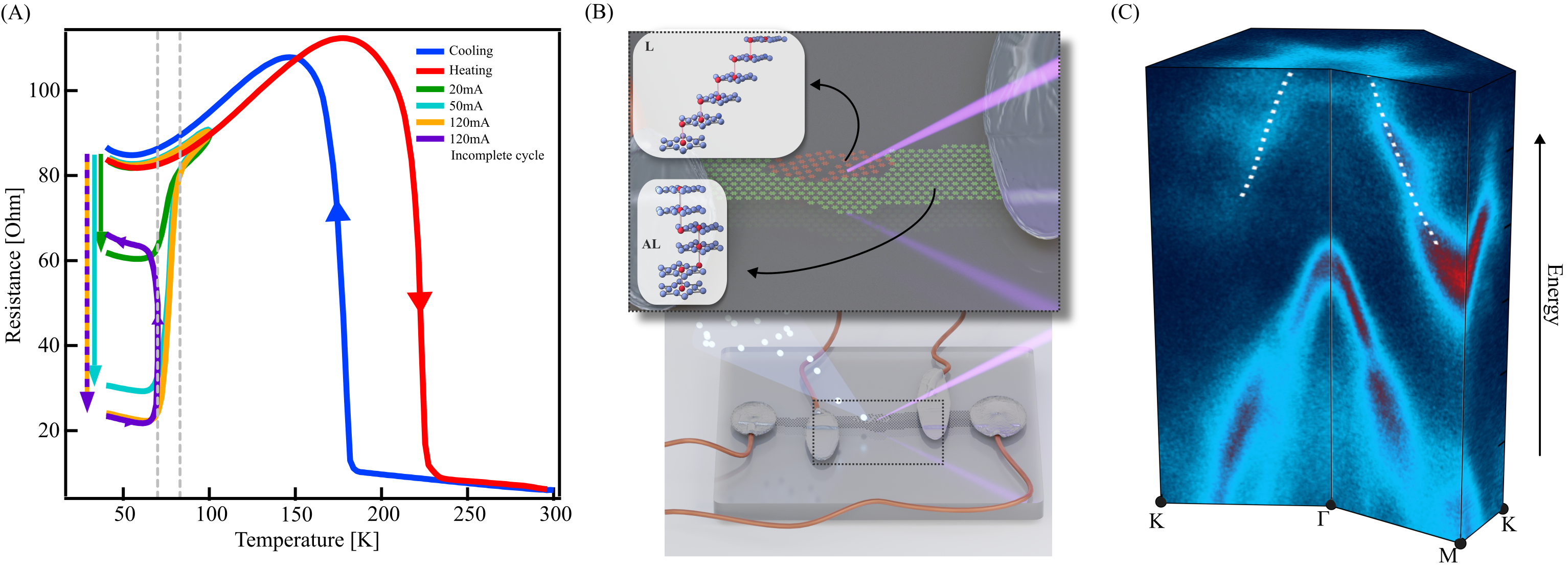}
    \caption{\textbf{The hidden meta-stable state in 1T-TaS$_2$.}  \textbf{(A)} Resistance as a function of temperature for a typical sample used in the ARPES experiment. The blue and red curves represent the sample resistance during cooling and heating, respectively, without any perturbation. The green, cyan, and orange curves illustrate the reduction in resistance with the application of progressively larger current pulses. The purple curve shows an incomplete recovery to the "normal" state after the application of a $120~m\,A$ pulse. \textbf{(B)} Schematic illustration of the microARPES experiment. A thin 1T-TaS$_2$ sample is fixed to a sapphire substrate using Ag-epoxy. Four contacts are used to apply current and measure the voltage across the sample. A micron-sized beam spot is employed to scan the electronic spectrum of the device. In the hidden state, the sample is phase-separated into insulating and metallic areas. The out-of-plane dispersion reveals that the unit cell in the insulating regions is twice as large as that in the metallic regions, corresponding to AL-stacking and L-stacking, respectively \cite{lee2019origin}. \textbf{(C)} ARPES data from the metallic part along high symmetry lines. Following the current pulse, a Fermi surface emerges in parts of the sample. The dispersion of the NC-CDW at 300K is indicated by the dotted white line.}
    \label{fig:fig1}
\end{figure}

 Narrow bridges were cut from a single crystal of 1T-TaS$_2$ and four contacts were made using Ag epoxy (for more details about the sample preparation see the Methods section). In Fig.~\ref{fig:fig1}A, we present the resistance as a function of temperature for a typical bridge. An abrupt change in resistance is observed when cooling into the C-CDW state at approximately $180~K$ (blue line) and when warming back up at approximately $220~K$ (red line).

At a temperature of $40~K$, a current pulse was applied, drastically lowering the sample's resistance. This low resistance is maintained as long as the temperature remains below 70 K, which is consistent with the transition temperature observed when a short light pulse was used to drive the transition~\cite{vaskivskyi_controlling_2015}.

As higher current pulses are applied, the sample's resistance decreases further, eventually saturating at around 140-190 mA for our typical sample dimensions. Upon warming, the resistance returns to its initial value. If the sample is cooled before a complete transition back to the C-CDW phase, it becomes locked in an intermediate state (see the purple line in Fig.~\ref{fig:fig1}A).

\subsection*{Spatial distribution of the hidden state}
\begin{figure}[h]
    \centering
    \includegraphics[width = 0.6\textwidth]{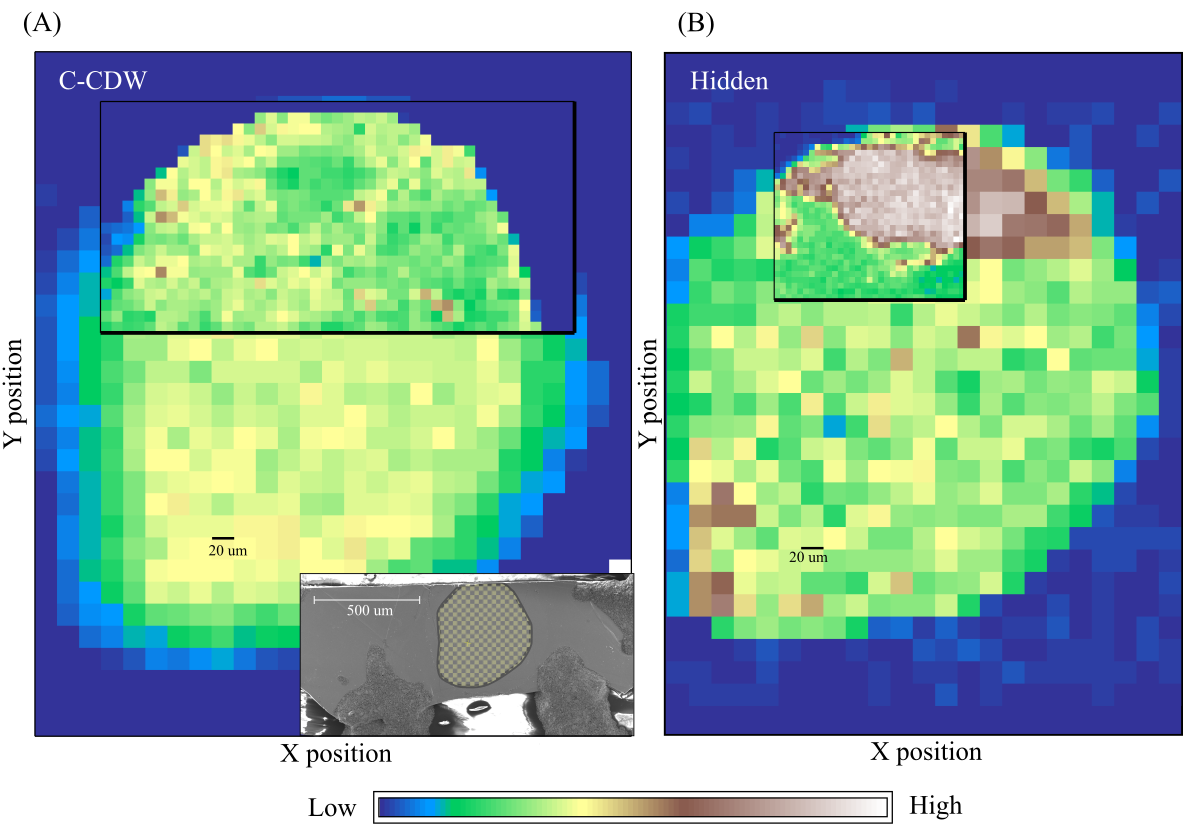}
    \caption{\textbf{Maps of the spectral weight at the Fermi level}. microARPES maps were measured with a spot size of 5 microns at 40 K. The step size was set to 20 microns, and was reduced in the area marked by a solid black line to 10 microns in (A) and 5 microns in (B). The color map represents the integrated ARPES intensity at the Fermi level.
In \textbf{(A)}, we show the sample in the C-CDW state. The sample was warmed and then cooled back down after being measured in the hidden state. The entire scan shows low intensity, indicating a gapped state. The inset is a SEM image showing the sample with Ag-epoxy contacts, with the cleaved area marked on the image.
In \textbf{(B)}, we show a spatial scan after a $190~mA$ current pulse. High-intensity regions can be seen. In these areas, the gap is closed, and we observe a metallic Fermi-Dirac edge.}
    \label{fig:new_spatial}
\end{figure}

In Fig. \ref{fig:new_spatial} we show  maps of the spectral weight at the Fermi-level over the entire cleaved area in the C-CDW state (Fig.~\ref{fig:new_spatial}A) and in the hidden state (Fig.~\ref{fig:new_spatial}B). The color map represents the momentum-integrated intensity of the microARPES spectra at the Fermi-level. The spatial resolution is set by the spot size which is about 5 microns in this case. The inset of Fig.~\ref{fig:new_spatial}A shows an SEM image of the measured sample. The cleaved area is marked.

In the C-CDW state, we observe a small, spatially uniform spectral weight at the Fermi level, originating from the tail of the flat band located at a binding energy of approximately 100 meV. In the hidden state, however, regions of high intensity indicate the metallic parts of the sample where the spectral gap is closed. The transition between the C-CDW and hidden states is fully reversible; in fact, the data in Fig.~\ref{fig:new_spatial}A was measured after the data in Fig.~\ref{fig:new_spatial}B by warming the sample and cooling it down again.

The use of a current pulse to drive the transition results in an inherently inhomogeneous state. Unlike a laser pulse, the current flow profile between the contacts cannot be precisely controlled and is generally non-uniform. Consequently, it is reasonable to assume that metallic regions form where the current density is sufficiently high. This inhomogeneity is not a characteristic of the hidden state but rather specific to the sample.

We observed different patterns of phase separation in the samples we measured and were unable to find defects on the sample surface that pin the metallic parts in the hidden state, suggesting that the switching is governed by the bulk current flow profile. Recently, it has been suggested that the metallic areas are more likely to form at the sample boundary~\cite{devidas2024spontaneous}.

\subsection*{Band dispersion of the hidden state}
\begin{figure}[h]
    \centering
    \includegraphics[width=0.8\textwidth]{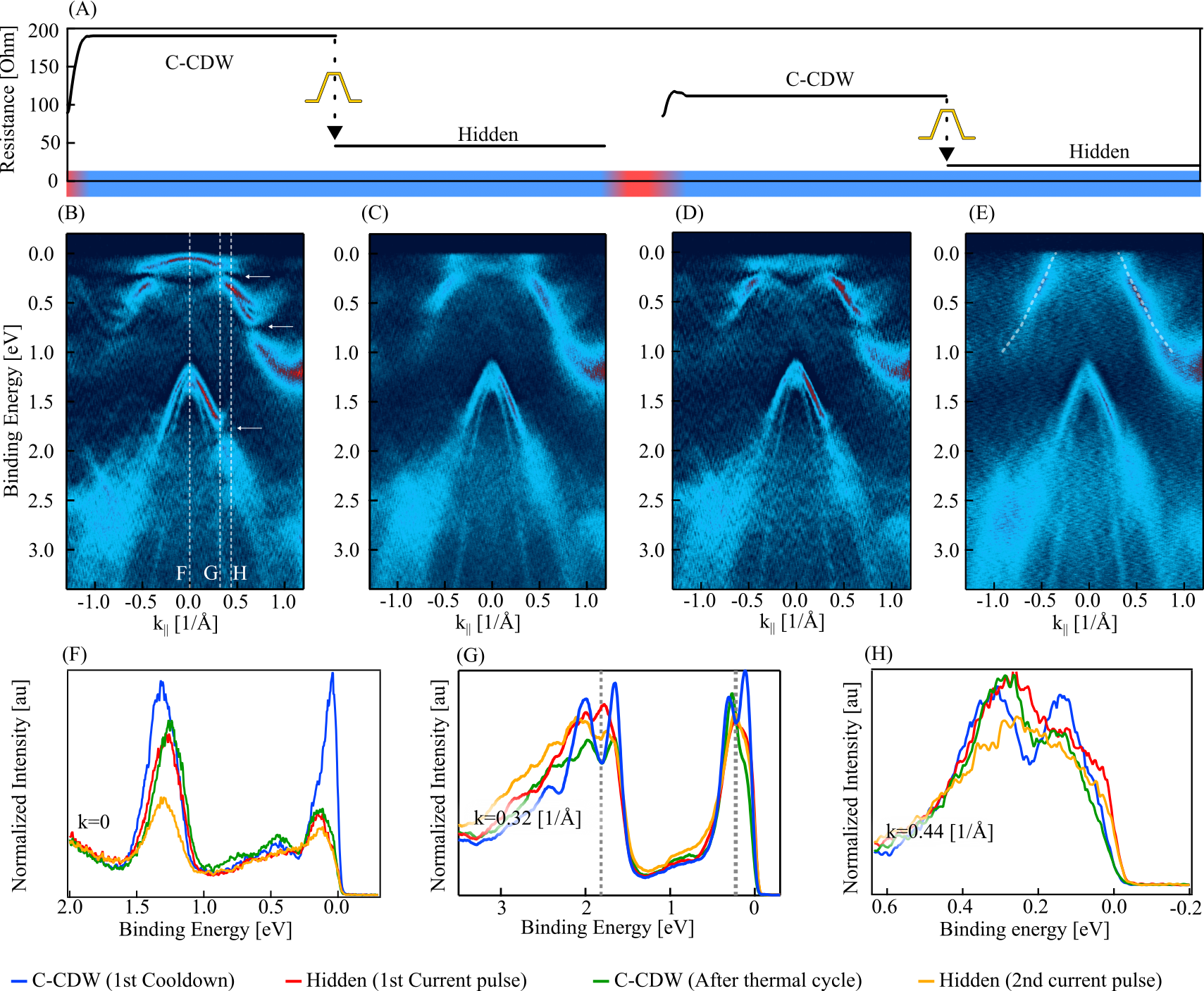}
    \caption{\textbf{ARPES spectra in the hidden state.} microARPES spectra at 40 K,  measured with 80~eV photons along the $\Gamma$-M direction. \textbf{(A)} Experimental timeline, showing the sample's resistance plotted as a function of time. The colorbar represents the temperature of the sample over time. 
    \textbf{(B)} ARPES image after the initial cool-down, in the C-CDW state. The resistance of the sample at this stage is $\sim200~\Omega$. One can see the characteristic flat-band at $\Gamma$ and the CDW gaps marked by white arrows. The dashed white lines indicate the momenta at which the EDCs are displayed in panels (F)–(H).  \textbf{(C)} the spectra after a $140~mA$, 1~ms long current pulse, the resistance dropped to $46~\Omega$. The CDW is highly suppressed, as evidenced by the suppressed band-folding and CDW gaps.  The band is crossing the Fermi level at $k_{\parallel}=0.44 \angstrom^{-1}$. \textbf{(D)} ARPES image after the sample was removed from the manipulator, allowed to warm up, and then reinserted and cooled down. The sample resistance is $115 \Omega$ , not fully restored. The CDW gaps are partially recovered. \textbf{(E)} ARPES image after a second, $140~mA$,  pulse was applied. The resistance dropped to $20 \Omega$. Again, the flat band and the CDW gaps vanish. The dashed white line shows the dispersion in the NC-CDW state measured at room temperature using standard ARPES. The measurements in (B)-(D) were all measured at approximately the same position on the sample. 
    \textbf{(F)} EDCs at the $\Gamma$ point . The flat band is suppressed by the pulses and only partially recovered in (C). The folded band due to the C-CDW is visible as a peak at $-450~meV$ in the C-CDW state(green,blue) while absent in the hidden state (red,orange). \textbf{(G)} EDCs at k$=0.32~\angstrom^{-1}$. Gray dotted lines show the C-CDW gaps position. The gap is clear in the C-CDW state (blue, green) and it is strongly suppressed in the hidden state (red, orange). \textbf{(H)} EDCs taken at k$_F=0.44~\angstrom^{-1}$ show a metallic Fermi-edge for the EDCs measured in the hidden state (red, orange).}
    \label{fig:pulse_progress_panel}
\end{figure}

Next, we compare in detail the spectra in the C-CDW state and the metallic regions of the hidden state. In Fig.~\ref{fig:pulse_progress_panel}, we present typical microARPES data along the $\Gamma-M$ direction, measured using $80~eV$ photon energy and a sub-micron spot size. Fig. \ref{fig:pulse_progress_panel}A shows the timeline of the experiment. First, the sample was cooled to $40~K$. The data before any current pulse was applied is shown in Fig.~\ref{fig:pulse_progress_panel}B, revealing the signature C-CDW gaps and a flat band around the $\Gamma$ point. A 140 mA, 1 ms long pulse was then applied, reducing the resistance by a factor of four. Fig.~\ref{fig:pulse_progress_panel}C presents the ARPES data after the current pulse, showing a suppression of the CDW gaps and a band crossing the Fermi level at k$_F=\pm0.4~\angstrom^{-1}$. The intensity of the flat band at the $\Gamma$ point is significantly suppressed. The sample was then removed from the cryo-manipulator and allowed to warm up in the vacuum chamber for approximately 10 minutes before being cooled back to base temperature. Following this thermal cycle, the resistance partially recovered to about half of the original value. It is important to note that the exact temperature during this warm-up sequence is unknown. The data in Fig.~\ref{fig:pulse_progress_panel}D show developed C-CDW gaps and a partial recovery of the flat band.
A second current pulse was then applied and the resistance dropped to a tenth of the original resistance. Fig.~\ref{fig:pulse_progress_panel}E shows ARPES data after the second current pulse. Again, the C-CDW gaps are closed and a metallic band is formed. For comparison, we show in white dashed-line the band-dispersion measured at room temperature, in the NC-CDW state, using a $\sim$100 microns spot size. The agreement with the hidden state dispersion is remarkable. 

In Fig.~\ref{fig:pulse_progress_panel}F-H we show energy distribution curves (EDCs)  at k$=0$ (F) , k$=0.32~\angstrom^{-1}$ (G) and at k$_{F}=0.44~\angstrom^{-1}$ (H), showing the flat band and the C-CDW gaps.
At the $\Gamma$ point (F) we can see a suppression of the flat band in the hidden state. The band does not fully recover after the thermal cycle. The small peak at $\approx -0.45eV$ is a signature of the folded band due to the C-CDW and is absent in the hidden phase. 
In (G) we find well developed gaps in the EDCs that correspond to the C-CDW state (marked with gray dashed lines), and a suppression of the gap in the EDCs measured in the hidden state. In (H) we show a metallic Fermi-edge in the EDCs taken in the hidden state and a gap at the Fermi-level for EDCs in the C-CDW state.

\subsection*{Chirality of the Fermi surface}
\begin{figure}[h]
    \centering
    \includegraphics[width =0.7\textwidth]{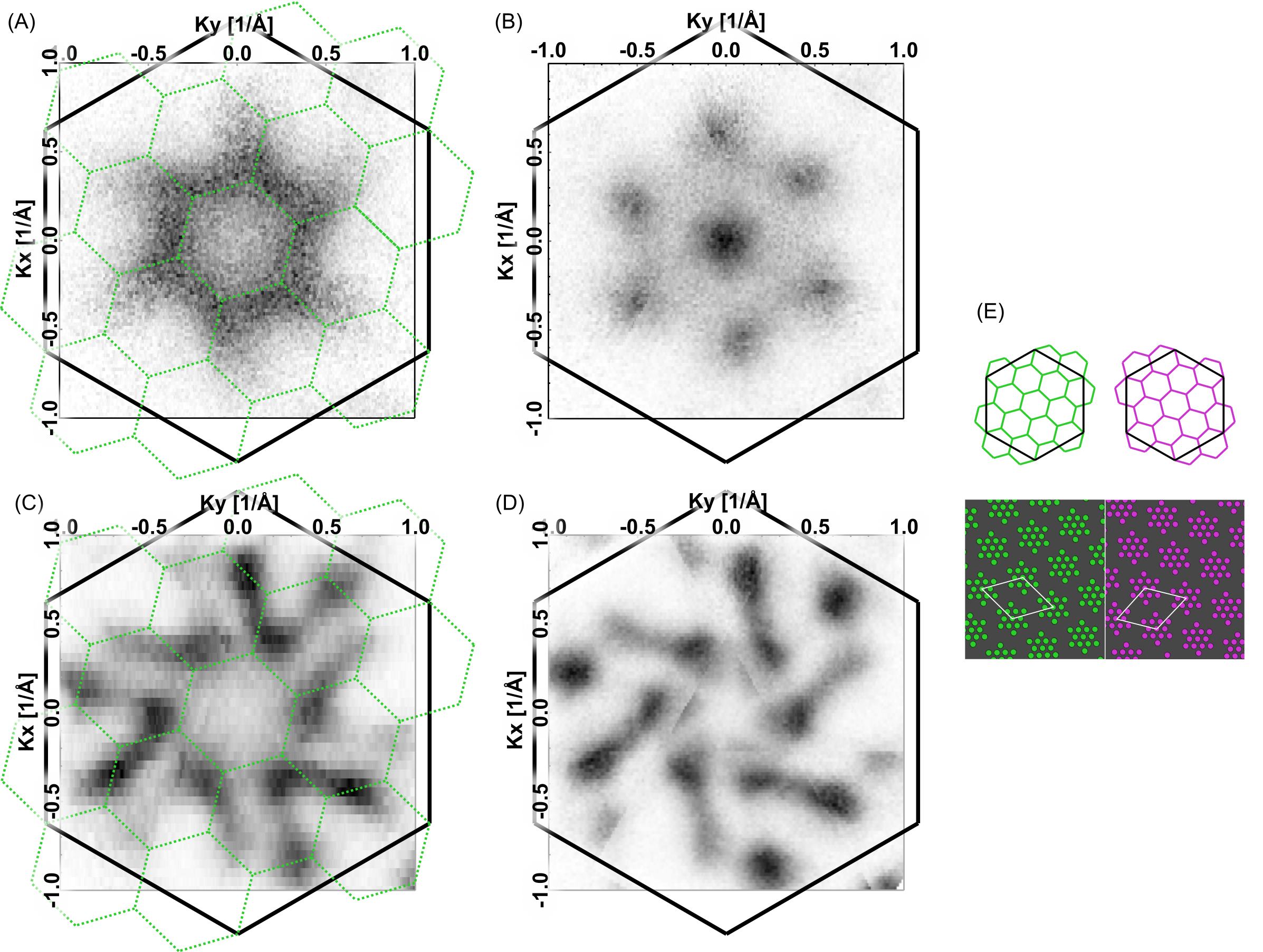}
    \caption{\textbf{ARPES intensity maps}. Constant energy maps at metallic and gapped regions in the hidden state. Data measured using 80~eV photon energy. The maps are generated by rotating the measured region by 120° to reconstruct the full Brillouin zone. The large hexagon in solid black line represents the normal state Brillouin zone (BZ). Small hexagons in green  are the CDW reconstructed BZs. \textbf{(A)} Fermi surface obtained from a metallic region by averaging the spectral weight within $\pm 10meV$ of the Fermi level. \textbf{(B)} Constant energy map from an insulating region,  measured at a binding energy of $30\pm 5~meV$. \textbf{(C)-(D)} Constant energy maps measured at $300 \pm5~meV$ in the metallic and insulating parts, respectively. Both show a distinct chirality which  originates from the formation of the star-of-David clusters. We find the same  chirality in the metallic and insulating regions. 
    \textbf{(E)} The two possible mirror-symmetry breaking C-CDW configurations. top: BZ of each configuration resulting from the real space C-CDW configuration illustrated below. }
    \label{fig:Fermi_surface}
\end{figure}

ARPES intensity maps measured in the hidden state at two different locations on the sample, one in a metallic region and the other in a gapped region are shown in Fig.~\ref{fig:Fermi_surface}.
In the metallic part, at the Fermi-level, we find a clear Fermi-surface (Fig.~\ref{fig:Fermi_surface}A). This FS is the origin of the enhanced conductivity in the hidden state. The FS is similar to the FS found in  the high temperature NC-CDW state~\cite{pillo1999remnant}. It consists of six elliptical pockets centered around the M-points. The Fermi pockets are gapped near the BZ edges, resulting in an incomplete Fermi surface~\cite{sohrt_how_2014,rossnagel_origin_2011}. 

In the gapped regions, there is no Fermi surface, but we observe intensity at low binding energies extending almost to the Fermi level. Fig.~\ref{fig:Fermi_surface}B shows an intensity map at a binding energy of $30~meV$. We find some spectral weight at the $\Gamma$ point of the atomic Brillouin zone (solid black lines) and weaker replicas at the $\Gamma$ points of the CDW-folded Brillouin zones (green dashed lines).

The electronic structure at higher binding energies indicates that a CDW deformation persists in the hidden state, even though the gaps are suppressed. In Fig.~\ref{fig:Fermi_surface}C and D, we compare intensity maps at a binding energy of 300 meV in a metallic region (C) and a gapped region (D). Both datasets exhibit a chiral pattern attributed to mirror symmetry breaking caused by the formation of star-of-David clusters~\cite{yang_chirality_2022}.
We emphasize that no measurable differences were observed between the gapped regions of the hidden state and the C-CDW spectra (see ~figure~\ref{fig:sup_CCDW_vs_gapped_hidden}).

Throughout the text, we refer to the band structure of 1T-TaS$_2$ as chiral, in accordance with the accepted notation in the field. However, it is important to note that 1T-TaS$_2$ is centrosymmetric, and therefore not chiral in the crystallographic sense. Instead, it is more accurately described as {\it pseudochiral}, a state characterized by the absence of mirror symmetry in the bulk 3D band structure~\cite{louat2024pseudochiral}.

The two possible configurations of the BZ in the C-CDW phase are illustrated in Fig.~\ref{fig:Fermi_surface}E. Remarkably, our results indicate that the resulting Fermi surface also breaks mirror symmetry, seemingly inheriting its chirality from the underlying CDW domains.

Moreover, by comparing the chiral patterns at the same metallic point both in the hidden state and prior to the application of the current pulse, we show that the chirality remains unchanged by the current pulse (see ~figure~\ref{fig:sup_unchanged_chirality}). This evidence, shown in the supplementary material, strongly suggests that the transition into the hidden state does not involve a complete melting of the CDW. This conclusion is in agreement with the distinct chirality of domains observed in real-space STM data~\cite{gerasimenko_intertwined_2019}.

The data from the metallic parts of the hidden state, both the dispersion and the Fermi surface, closely resemble those of the NC-CDW stat~\cite{lahoud2014emergence}. However, STM measurements reveal that in the hidden phase, the C-CDW is fragmented into a disordered network of domains, each approximately tens of nanometers in size~\cite{gerasimenko_intertwined_2019}. This domain configuration is distinct from that observed in the NC-CDW state.
The spot size used in our experiment covers approximately 1000 domains. Given that the domain walls are around $\sim1\angstrom$ in width~\cite{wu1989hexagonal}, their contribution to the ARPES spectra is negligible. Combined with the distinct chirality of the Fermi surface we find(see ~figure~\ref{fig:sup_Hidden_FS_chirality}), this confirms that the metallic nature of the hidden state is not due to conducting domain walls.

\subsection*{Out-of-plane dispersion measurements}
\begin{figure}[h]
    \centering
    \includegraphics[width=0.7\textwidth]{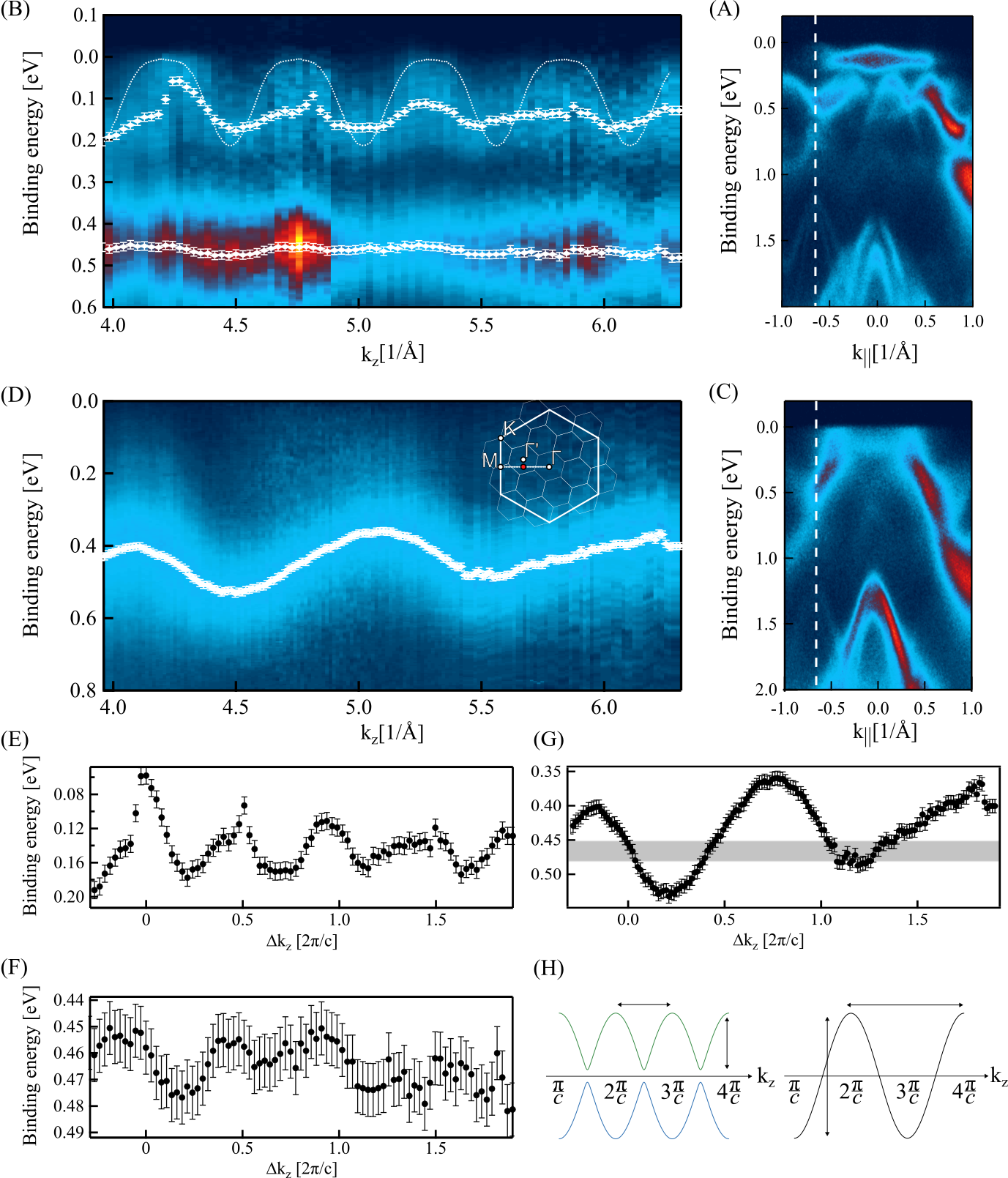}
    \caption{\textbf{Out-of-plane dispersion}. \textbf{(A)} ARPES image along the $\Gamma$-M  direction from a gapped region in the hidden state measured with $77~eV$ photons. White dotted line marks k$_\parallel$ where the photon energy scan was measured. \textbf{(B)} Intensity as a function of binding energy and k$_z$ for photon energies between $48~eV$ and $140~eV$. White diamonds indicate the binding energy of the spectral peak at each $k_z$, extracted from fits to the EDCs; error bars correspond to the larger of the fit uncertainty and the experimental energy resolution ($10~\mathrm{meV}$). The dashed white line represents the calculated out-of-plane dispersion in the AL-stacking configuration from Ref.~\cite{Stacking-driven_gap_Ritschel_2018}. Each EDC is normalized by the intensity at the peak at low binding energies. The dispersion of the two bands exhibits a periodicity that is double the expected value based on the atomic unit cell size. \textbf{(C)} and \textbf{(D)} same as (A) and (B) measured in a metallic region. Here, the dispersion periodicity follows the atomic BZ. Inset: the red dot marks the location in k$_\parallel$ at which the k$_z$ data was measured. It is close to the $\Gamma'$ point. \textbf{(E)} and \textbf{(F)} the k$_z$ dispersion of the flat band and the second valance band, respectively, as extracted from panel (B). \textbf{(G)} the k$_z$ dispersion of the top most band in the metallic region as extracted
from (C). The shaded gray area represents the bandwidth of the band shown in (F). \textbf{(H)} Schematic illustration of the k$_z$  dispersion. The right side shows the dispersion of a band with 1 electron per unit cell. The left side shows the dispersion when the unit cell is doubled, either due to dimerization or a 2-unit cell CDW}
    \label{fig:kz_CDW}
\end{figure}

After confirming that the CDW persists in both the metallic and gapped regions of the sample following the current pulse, we turn our attention to understanding the differences between these regions.  In Fig. \ref{fig:kz_CDW},  we present photon-energy dependent data measured at representative points in both areas. The out-of-plane dispersion was recorded at a fixed parallel momentum,  
k$_\parallel=-0.66\angstrom^{-1}$ (indicated by the red dot in the inset of Fig. \ref{fig:kz_CDW}D). This point lies near the center of the reconstructed Brillouin zone, $\Gamma^{'}$, as seen by the faint replica of the band at a binding energy of $\sim 1.25~eV$ . 

Panels \ref{fig:kz_CDW}(A and B) correspond to a gapped region while panels \ref{fig:kz_CDW}C and D are from a metallic region.  
 Fig. \ref{fig:kz_CDW}A and C show data along $\Gamma$-M measured using 77 eV photons in the insulating and metallic regions, respectively. The white dashed lines represent the in-plane momentum at which the dispersion out-of-plane was measured. In Fig. \ref{fig:kz_CDW}B and D we show a color plot combining the EDCs at different k$_z$ values.

 The k$_z$ dispersion of the two uppermost sub-bands formed by the C-CDW, extracted by tracking the maxima in the EDCs, are shown in Fig. \ref{fig:kz_CDW}E and F. 

    The bandwidths of the two bands are approximately 100 meV and 50 meV, respectively. Notably, both bands show a clear doubling of the expected periodicity, indicating a doubling of the unit cell. This observation is consistent with a charge density wave out-of-plane with two unit-cell periodicity or a dimerization of pairs of star-of-David clusters on two neighboring planes.
The CDW replicas exhibit significantly weaker intensity compared to the main band. The double periodicity becomes more pronounced when the dispersion is measured at the $\Gamma'$ point instead of the $\Gamma$ point. The dashed white line in Fig.\ref{fig:kz_CDW}B represents the results of a DFT calculation for an alternating stacking t$_{02}$ configuration, where two vertically aligned star-of-David clusters form pairs within each bilayer, and these pairs are shifted relative to the pairs in neighboring bilayers \cite{Stacking-driven_gap_Ritschel_2018}.  Our results show qualitative agreement with the DFT calculation.

We observe a clear doubling of the unit cell along the out-of-plane direction, with the spectrum gapped at all k$_z$ values. Previous studies were somewhat inconclusive on this point \cite{Hoffman_kz,PhysRevB.106.155406}. The difference could be related to the significantly smaller light spot used in our study.

When examining the k$_z$ dispersion in the metallic regions, we observe a different behavior. In Figs. \ref{fig:kz_CDW}C and \ref{fig:kz_CDW}D, we present data along $\Gamma$-M measured using $77~eV$ photons, and photon energy dependence of the EDCs measured in a metallic part of the sample, respectivley. The extracted dispersion is shown in Fig. \ref{fig:kz_CDW}G, revealing only one band at this k$_\parallel$ value. The bandwidth is approximately 200 meV, which is significantly larger than that observed in the insulating region. Notably, we do not find the double periodicity; instead, the dispersion is in agreement with the atomic lattice unit cell.

\section*{Discussion}
Our study reveals that the hidden state in 1T-TaS$_2$ hosts a hole-like band crossing the Fermi level, resulting in a Fermi surface reminiscent of the nearly commensurate NC-CDW phase. This similarity is not surprising, given the structural resemblance between the two phases—both characterized by domains of commensurate CDW order. However, the mechanisms underlying the metallicity of these phases remain elusive. While it has been proposed that conducting domain walls—regions where the order parameter $\Delta = 0$ - form channels that contribute to the overall conductivity, our ARPES measurements suggest that this mechanism alone cannot account for the observed metallic behavior. Although such channels may be locally conductive, they are insufficient to explain the emergence of a robust Fermi surface and substantial spectral weight at the Fermi level.

We show that the hidden Fermi surface exhibits distinct chirality arising from the C-CDW domain arrangements - an intrinsic property of the domains themselves rather than the domain walls. Furthermore, the significant spectral weight observed in ARPES, which is surface-sensitive, indicates that each domain is intrinsically metallic, and the contribution from domain walls is minimal. Importantly, the in-plane CDW motif remains essentially intact across both insulating and metallic regions, suggesting that the metallicity does not result from CDW melting or disordering, but from a more subtle structural transformation.

Our findings establish a compelling link between the stacking configuration of 1T-TaS$_2$ and its emergent electronic properties. While the monolayer is likely a Mott insulator \cite{vavno2021artificial}, in the bulk, first-principles calculations have shown that the out-of-plane stacking of star-of-David clusters plays a decisive role in shaping the in-plane electronic dispersion, modulating gap structures and bandwidths depending on the stacking configuration \cite{lee2019origin,ritschel2018stacking}. Specifically, DFT calculations by Lee et al. \cite{lee2019origin} indicate that the AL-stacking configuration is energetically the most favorable, with the L-stacking only slightly higher in energy. This small energy difference suggests that a stacking rearrangement from AL to L can be induced by a moderate external perturbation such as a current pulse. 

To uncover the nature of this transformation, we examined the out-of-plane band dispersion at binding energies where differences between metallic and insulating regions are minimized, isolating the influence of interlayer effects. In the insulating regions, the k$_z$ dispersion shows a doubled periodicity and reduced bandwidth, consistent with interlayer dimerization and a doubling of the unit cell along the $c$-axis, as previously reported \cite{stahl_collapse_2020}. These features align with the AL-stacking configuration, in which vertical dimers of star-of-David clusters are stacked such that the central atom of one dimer aligns with one of the outer atoms of a neighboring cluster (See Fig. \ref{fig:fig1}B). The dashed white line in Fig.\ref{fig:kz_CDW}B shows the calculated out-of-plane dispersion for this configuration, which matches well with our data from the insulating regions.

In contrast, the metallic regions exhibit a single-period k$_z$ dispersion closely matching the primitive lattice periodicity and a broader bandwidth, indicating the absence of dimerization, in agreement with the DFT predictions for the L-configuration (See Fig. \ref{fig:fig1}B). These observations suggest that the current pulse disrupts the AL-stacked (dimerized) configuration, breaking the interlayer correlations that stabilize the insulating state. The system then transitions into the L-stacking configuration, a nearly degenerate configuration in energy, which restores a simpler stacking periodicity and enables coherent interlayer hopping, thereby giving rise to a three-dimensional metallic state.

This hidden metallic phase represents an unusual electronic structure among transition metal dichalcogenides: the out-of-plane (interlayer) bandwidth exceeds the in-plane bandwidth. This is in agreement with transport measurements showing that the resistivity anisotropy in the NC-CDW state is close to one \cite{martino2020preferential}.

Such an inversion of dimensionality, where transport and electronic coherence are dominated by the out-of-plane direction, challenges conventional views of quasi-2D physics and may open new avenues for engineering anisotropic transport and manipulating quantum phases in layered materials.


\section*{Materials and Methods}
\subsection*{Sample preparation}
1T-TaS$_2$ single crystal were grown as described in Ref.~\cite{PhysRevB.96.195131}.
Samples of 1T-TaS$_2$ were prepared for nanoARPES experiments: Single crystals were cut into narrow bridges  200-500$~\mu\,m$ in width. The samples were glued to a sapphire substrate and 4 contacts were made using Ag epoxy. The samples were cooled to $40~K$ and cleaved under UHV conditions. The 4 contacts were used to drive a current pulse and to monitor the sample resistance in-situ. We emphasize that during the ARPES measurement no current was applied to the sample.

\subsection*{ARPES measurements}
Spatially resolved ARPES experiments were carried out in the I05 beamline at Diamond light source, UK. The spatial resolution is set by the beam spot-size, which in our experiment was either ~500~nm using a Zone plate or 5 micrometers using a capillary mirror. Energy resolution was set to about 50~m\,eV.
Data in the NC-CDW state was measured at the CASSIOPEE beamline at Soleil, France. Measurements were done at room temperature using 80~eV photon energy. Beam spot size was ~100~microns. Energy resolution was set to 20~m\,eV. 
The Fermi-level was determined by measuring a metallic sample at the same conditions. 

\subsection*{calculation of k$_\perp$}
We have used the three-step-model for the out-of-plane momentum k$_z=\sqrt{\frac{2m}{\hbar^2}(E_k\cos^2\theta+V_0)}$~\cite{damascelli2004probing}, where $V_0$ is the inner potential. We have used $V_0\sim15.4~eV$, obtained by following the normal-crystal periodicity over 3 BZs, demanding $\Delta k_z^\Gamma = \frac{2\pi}{c}$ ($c=5.861~ \angstrom$).
\subsection*{Band fitting}
Peak fitting was done using Gaussian functions. When close to $E_F$, the Gauassians were multiplied by the Fermi function, where $E_F$ was fixed to zero, and the temperature was set to $\approx 20~K$.

\newpage


\clearpage 

%
\bibliography{hidden_state_TaS2} 
\bibliographystyle{sciencemag}

%
%
%
%
%
%


\section*{Acknowledgments}
We thank Assa Auerbach and Daniel Podolsky for useful discussions.
We acknowledge Diamond Light Source for time on Beamline i05 under Proposal SI33131.
We acknowledge SOLEIL for provision of synchrotron radiation facilities under proposal 20230771 and we would like to thank  Chiara Bigi and Francois Bertran for assistance in using beamline CASSIOPEE.\\
\textbf{Funding:}
The work at the Technion was supported by Israeli Science Foundation grant number ISF-1263/21.\\
\textbf{Author contributions:}
A.K. conceived the experiment. I.F. prepared the 1T-TaS2 single crystals. Y.N. performed the ARPES experiments and analyzed the ARPES data. R.A.G.K, I.M., A.D., N.R. S.K.P. and Y.J. performed ARPES measurements. A.L., M.D.W. and C.C. provided support at the synchrotron beam-line. Y.N. and A.K. wrote the manuscript with contributions from all authors.
\textbf{Competing interests:}
The authors declare that they have no competing interests.\\
\textbf{Data availability}:
All data needed to evaluate the conclusions in the paper are present in the paper and/or the Supplementary Materials.

\subsection*{Supplementary materials}





\renewcommand{\thefigure}{S\arabic{figure}}
\renewcommand{\thetable}{S\arabic{table}}
\renewcommand{\theequation}{S\arabic{equation}}
\renewcommand{\thepage}{S\arabic{page}}
\setcounter{figure}{0}
\setcounter{table}{0}
\setcounter{equation}{0}
\setcounter{page}{1} 





Figures S1 to S3\\
\newpage



\newpage


\begin{figure}
    \centering
    \includegraphics[width=0.6\textwidth]{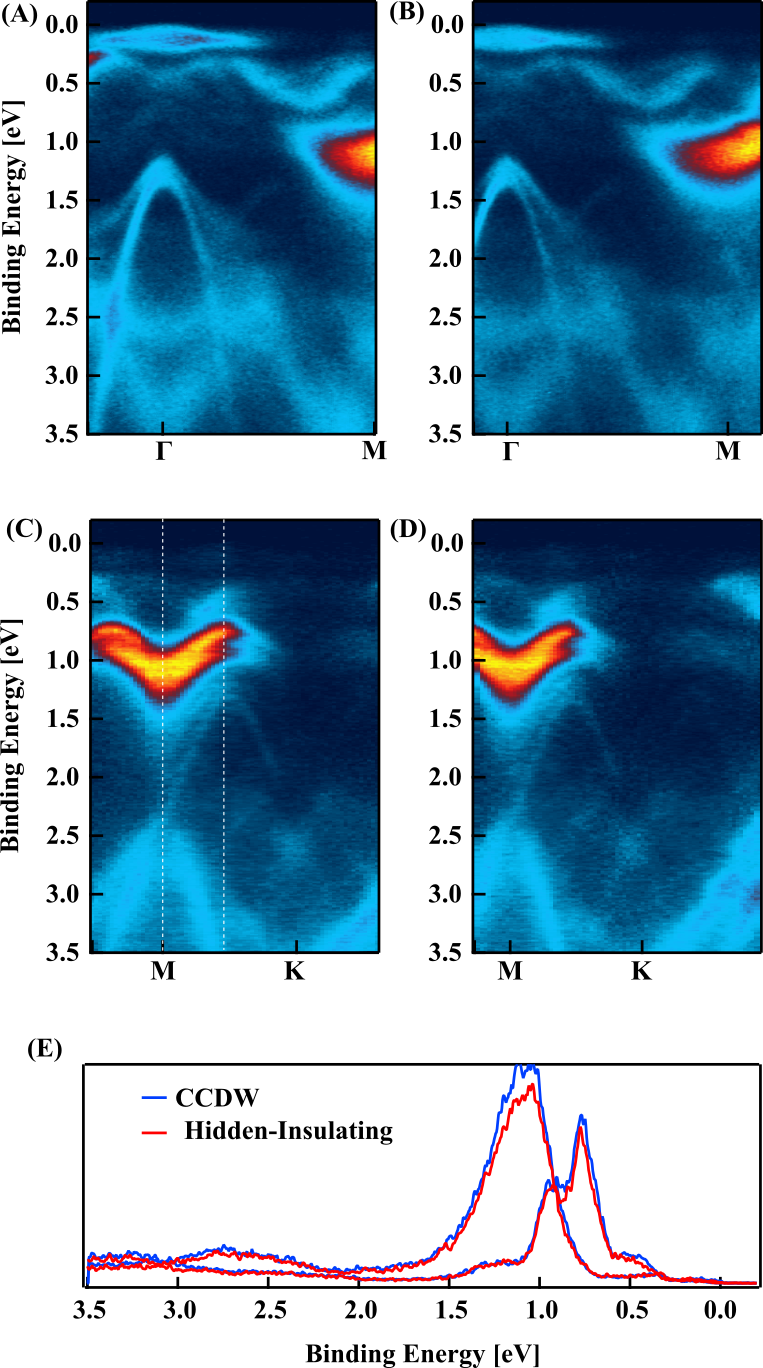}
    \caption{\textbf{Comparing the spectrum in the C-CDW state and in the gapped parts in the hidden state } \textbf{(A-B)} ARPES images along the $\Gamma$-M direction for the C-CDW state and the gapped part of the hidden state, respectively. \textbf{(C-D)} ARPES images along the M-K direction for the C-CDW state and the gapped part of the hidden state, respectively. \textbf{(E)} EDCs at the momenta indicated by the dotted lines in (C), for both the C-CDW state and the gapped part of the hidden state.}
    \label{fig:sup_CCDW_vs_gapped_hidden}
\end{figure}

\begin{figure}
    \centering
    \includegraphics[width=0.9\textwidth]{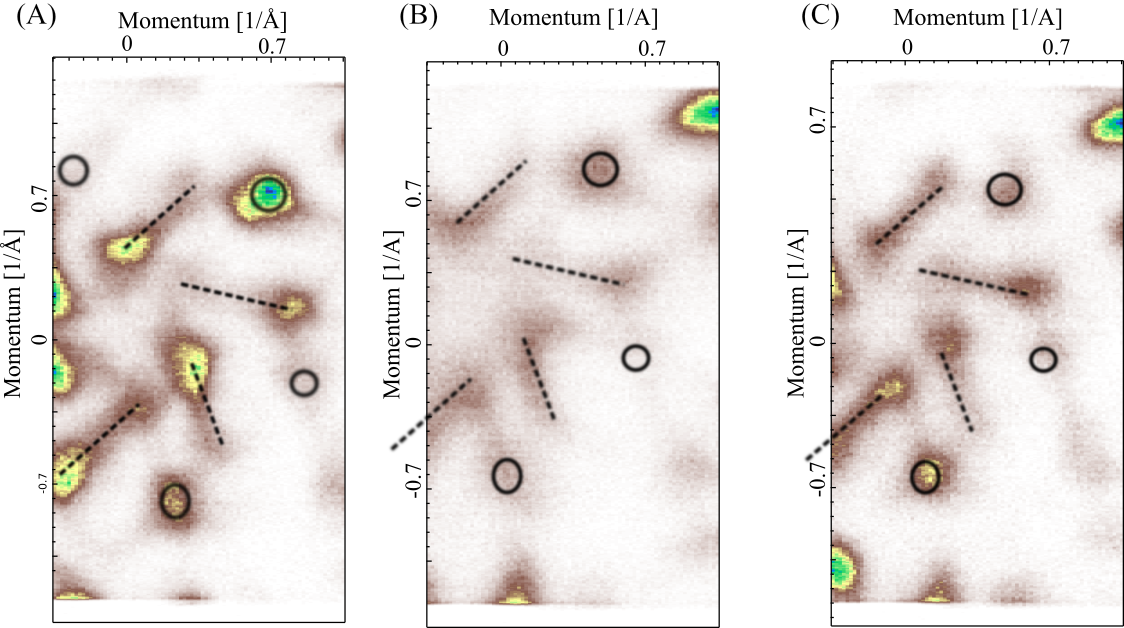}
    \caption{\textbf{Polar maps at the same location on the sample.} ARPES polar maps measured with 72~eV photon energy, at $E_b=300~m\,eV$ \textbf{(A)} after initial cooldown \textbf{(B)} after pulse  \textbf{(C)} after heating and cooling again (approximately at the same location).}
    \label{fig:sup_unchanged_chirality}
\end{figure}

\begin{figure}
    \centering
    \includegraphics[width=0.9\textwidth]{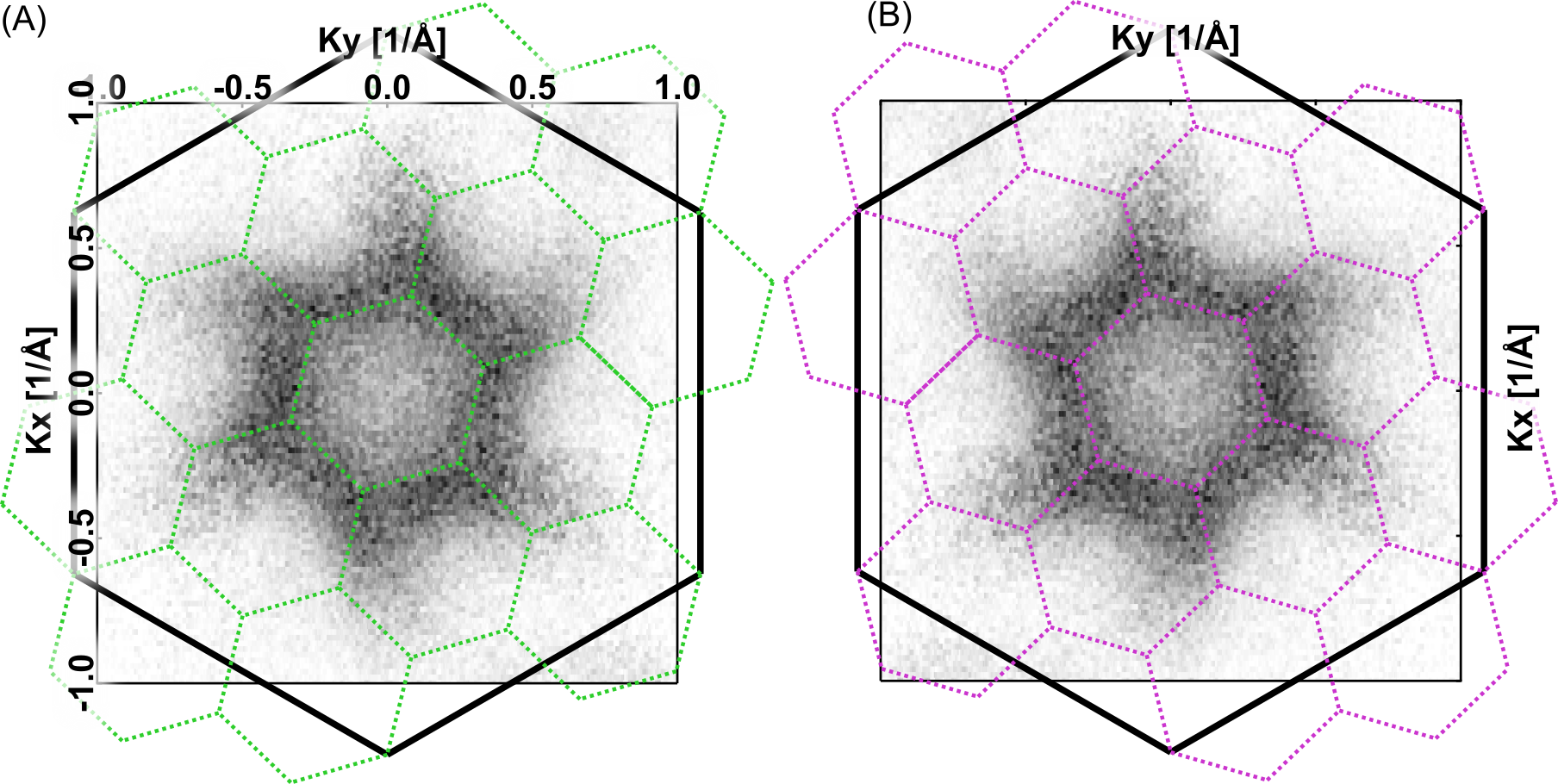}
    \caption{\textbf{Hidden state Fermi-surface chirality} \textbf{(A)} FS of the hidden state, as shown in \ref{fig:Fermi_surface} \textbf{(B)} same as (A) but horizontally mirrored }
    \label{fig:sup_Hidden_FS_chirality}
\end{figure}





\end{document}